\documentclass[12pt]{revtex4}
\usepackage{mathrsfs} %
\usepackage{dcolumn}
\usepackage{bm,amssymb,amsmath}

\begin{document}
\title{\textsf{Neutrino oscillations in Kerr-Newman space-time}}
\author{Jun Ren$^{1}$\footnote{Email:
renjun@hebut.edu.cn} and Cheng-Min Zhang$^{2}$}

\address{1 School of Science, Hebei
University of Technology, 300130, Tianjin, China
\\2 National
Astronomical Observatories, Chinese Academy of Sciences, 100012
Beijing, China}
\date{\today }

\begin{abstract}
The mass neutrino oscillation in Kerr-Newman(K-N) space-time is
studied in the plane $\theta=\theta_{0}$, and the general equations
of oscillation phases are given. The effect of the rotation and
electric charge on the phase is presented. Then, we consider three
special cases: (1) The neutrinos travel along the geodesics with the
angular momentum $L=aE$ in the equatorial plane. (2) The neutrinos
travel along the geodesics with $L=0$ in the equatorial plane. (3)
The neutrinos travel along the radial geodesics at the direction
$\theta=0$. At last, we calculate the proper oscillation length in
the K-N space time. The effect of the gravitational field on the
oscillation length is embodied in the gravitational red shift
factor. When the neutrino travels out of the gravitational field,
the blue shift of the oscillation length takes place. We discussed
the variation of the oscillation length influenced by the
gravitational field strength, the rotation $a^{2}$ and charge $Q$.


\end{abstract}

\maketitle

\section{Introduction}
 Mass neutrino mixing and oscillations were proposed by
Pontecorvo\cite{pon}, and Mikheyev, Smirnov and Wolfenstein
 (MSW for short) explored the effect of transformation of one neutrino
 flavor  into another  in a medium with varying
density\cite{Mikheyev,wolf}. Recently, the consideration of the mass
neutrino oscillations has been a hot topic. There have been many
theoretical\cite{ponte,bilenky2,nuss,kay,giunti3,rich,kiers,mann4}
and experimental\cite{fuk,apo,egu,ahn,ahm,mic,aha} studies about the
neutrino oscillations. Then, the neutrino oscillations in the flat
space time were extended to the cases in the curved
space-time\cite{ahlu,pir,null,fuller,zhangcm,zhangcm4,pereira,wang1}.
Neutrino oscillation experiments were also considered to test the
equivalence principle\cite{mann3}. Calculating the phase along the
geodesic line will produce a factor 2 in the high energy limit,
compared with the value along the null line, which often exists in
the flat and Schwarzschild
space-time\cite{zhangcm,zhangcm4,12,lipkin,lipkin2,grossman}. This
issue of the factor 2 is due to the difference between the time-like
and null geodesics. Furthermore, some alternative mechanisms have
been proposed to account for the gravitational effect on the
neutrino oscillation\cite{alter,mann1,mann2}. The inertial effects
on neutrino oscillations and neutrino oscillations in non-inertial
frames were also called attention\cite{lamb1,lamb2}. As a further
theoretical exploration, neutrino oscillations in space-time with
both curvature and torsion\cite{19,20,21} have been studied.

In recent years, the researches about the neutrino oscillation have
been made new progress. A further mechanism to generate pulsar
kicks, which was based on the spin flavor conversion of neutrinos
propagating in a gravitational field, and the neutrino geometrical
optics in gravitational field (in particular in a Lense-Thirring
background), have been proposed by Lambiase\cite{lamb3,lamb4}. Some
publications were centered on the theoretical study and experimental
measurement of the mixing angle $\theta_{13}$\cite{lyc,rat,eby}. And
CP violation in neutrino oscillations were considered by some
authors\cite{kli,sch,gav,alt}. In addition, Cuesta and Lambiase
studied the neutrino mass spectrum\cite{lamb}. Akhmedov, Maltoni and
Smirnov presented the neutrino oscillograms for different
oscillation channels and discussed the effects of non-vanishing 1-2
mixing\cite{smir}.

In this paper, we extend the mass neutrino oscillation work from
Schwarzschild space-time to Kerr-Newman space-time, since the
Kerr-Newman metric is rather important in black hole physics, where
a most generally stationary solution with axial symmetry has been
existing\cite{ruffini}. For the reason of simplicity, we confine our
treatment in two generation neutrinos (electron and muon). We give
the general equations of the oscillation phases along the arbitrary
null and the time-like geodesics, respectively, in the equal
$\theta$ plane, $\theta=\theta_{0}$. The phase along the geodesic
will also produce a factor of $2$ in the K-N space-time,
$\Phi(geod)=2\Phi(null)$, in the high energy limit. In our
derivation we have not assumed a weak field approximation.

We discuss three spacial cases. Firstly, the oscillation phases
along the geodesics with $L=aE$ are considered in the equatorial
plane. $E$ is the energy per unit mass of the particle. $L$ and $a$
are the angular momentum per unit mass of the particle and K-N
space-time, respectively. The geodesics with $L=aE$ in K-N
space-time play the same roles as the radial geodesics in the
Schwarzschild and in the Reissner-Nordstrom geometry. In this case,
the phases both along the null geodesic and the time-like geodesic
are similar in form to the phases along the radial geodesics (null
and time-like) in the Schwarzschild space-time. Secondly, we
calculate the oscillation phases along the geodesics with $L=0$ in
the equatorial plane. This kind of geodesics is also important in
K-N space-time. In the Schwarzschild space-time with non-rotating
spherically symmetry, particles with $L=0$ can propagate along the
radial geodesics. But in K-N space-time, because of dragging effect,
the coordinate $\varphi$ must change if a particle with $L=0$
travels along the geodesics. Thirdly, the phases along the radial
geodesics at the direction $\theta=0$ are given. Only at the poles
$\theta=0$ and $\theta=\pi$, the ergosphere coincides with the event
horizon. At the direction $\theta=0$, the effects of the rotation of
the space-time on the oscillation length are found to be more than
those in the other directions.

At last, we calculate the proper oscillation length in the K-N space
time. The oscillation length is proportional to the local energy
(local measurement), $E^{loc}=E/\sqrt{g_{00}}$, of the neutrino,
where $E$ is a constant along the geodesic. The decrease in the
local energy leads to the decrease in the oscillation length as the
neutrino travels out of the gravitational field. So, the blue shift
of the oscillation length occurs, which is unlike  the case of the
gravitational red shift for light signal. In the equatorial plane in
K-N space-time, the rotation have no contribution to the oscillation
length because $g_{00}$ has nothing to do with the rotating
parameter $a$ in this plane. The rotation $a^{2}$ of the
gravitational field shortens the oscillation length in other equal
$\theta$ plane, compared with the length in R-N space time. We also
give that the length varies according to $\theta$. And charge $Q$
shortens it too, compared with the Kerr space-time case. But, the
gravitational field lengthens it, compared with the case in flat
space-time.

In this paper, we take the neutrino as a spin-less particle to go
along the geodesic because the spin and the curvature coupling has a
little contribution to the geodesic derivation\cite{aud}. Moreover,
the neutrino is a high energy particle, so we do not think the
neutrino spin has more contribution to the geodesic.

The paper is organized as the follows. In Sec.2, we briefly review
the standard treatment of neutrino oscillation in the flat
space-time. In Sec.3, we give the general expressions of the
oscillation phases along the null and time-like geodesics in
arbitrary equal $\theta=\theta_{0}$ plane. In Sec.4, we discuss the
neutrino phase in three special cases. In Sec.5, we discuss the
proper oscillation length in K-N space-time. At last, the conclusion
and discussion are given. Throughout the paper, the units $
G=c=\hbar =1$ and $\eta_{\mu\nu}=diag(+1,-1,-1,-1)$ are used.
\section{the standard treatment of neutrino oscillation in flat space-time}
 In a standard treatment, the flavor eigenstate $\mid\nu_{\alpha}\rangle$
is a superposition of the mass eigenstates $\mid\nu_{k}\rangle$,
i.e.\cite{null,fuller}
\begin{equation}|\nu_{\alpha}\rangle=\sum_{k}U_{\alpha k}exp[-i\Phi_{k}]|\nu_{k}\rangle,\end{equation}
where\begin{equation}\Phi_{k}=E_{k}t-\vec{p}_{k}\cdot\vec{x},(k=1,2),\label{2}\end{equation}and
the matrix elements $U_{\alpha k}$ comprise the transformation
between the flavor and mass bases. $E_{k}$ and $\vec{p}_{k}$ are the
energy and momentum of the mass eigenstates $\mid\nu_{k}\rangle$,
and they are different for different mass eigenstates. If the
neutrino produced at a space-time point $A(t_{A},\vec{x}_{A})$ and
detected at $B(t_{B},\vec{x}_{B})$, the expression for the phase in
Eq.(\ref{2}), which is coordinate independent and suitable for
application in a curved space-time,
is\cite{null,stodolsky}\begin{equation}\Phi_{k}=\int_{A}^{B}p^{(k)}_{\mu}dx^{\mu},\end{equation}
where
\begin{equation}p^{(k)}_{\mu}=m_{k}g_{\mu\nu}\frac{dx^{\nu}}{ds}\label{p},\end{equation}is the canonical
conjugate momentum to the coordinate $x^{\mu}$ and $m_{k}$ is the
rest mass corresponding to mass eigenstate $|\nu_{k}\rangle$.
$g_{\mu\nu}$ and $s$ are metric tensor and an affine parameter,
respectively.

 The following assumptions are often applied in the standard
treatment\cite{ponte}: (1) The mass eigenstates are taken to be the
energy eigenstates, with a common energy; (2) up to $O(m/E)$, there
is the approximation $E\gg m$; (3) a massless trajectory is assumed,
which means that the neutrino travels along the null trajectory. In
the case of two neutrinos mixing $\nu_{e}-\nu_{\mu}$, we can
write\begin{equation}\nu_{e}=cos\theta\nu_{1}+sin\theta\nu_{2},\nu_{\mu}=-sin\theta\nu_{1}+cos\theta\nu_{2}.\end{equation}
Here $\theta$ is the vacuum mixing angle. The oscillation
probability that the neutrino produced as $|\nu_{e}\rangle$ is
detected as $|\nu_{\mu}\rangle$
is\cite{boehm}\begin{equation}\emph{P}(\nu_{e}\rightarrow\nu_{\mu})=|\langle\nu_{e}|\nu_{\mu}(x,t)\rangle|^{2}=sin^{2}(2\theta)sin^{2}(\frac{\Phi_{kj}}{2}),\end{equation}
where, $\Phi_{kj}=\Phi_{k}-\Phi_{j}$, is the phase shift. From the
standard treatment of the neutrino
oscillation\cite{null,fuller,zhangcm}, the standard result for the
phase
is\begin{equation}\Phi_{k}\simeq\frac{m^{2}_{k}}{2E_{0}}|\vec{x}_{B}-\vec{x}_{A}|.\end{equation}
Here $E_{0}$ is the energy for a massless neutrino. So, the phase
shift responsible for the oscillation is given by
\begin{equation}\Phi_{kj}\simeq\frac{\Delta m^{2}_{kj}}{2E_{0}}|\vec{x}_{B}-\vec{x}_{A}|,\end{equation}
where $\Delta m^{2}_{kj}=m^{2}_{k}-m^{2}_{j}$.
\section{neutrino oscillation phase along the null and the time-like geodesic in the plane $\theta=\theta_{0}$}

In this section, we study the neutrino oscillation in equal
$\theta=\theta_{0}$ surface. The line element of K-N space time
takes the form
\begin{equation}
ds^{2}=g_{00}dt^{2}+g_{11}dr^{2}+g_{22}d\theta^{2}+g_{33}d\varphi^{2}+2g_{03}dtd\varphi
\label{line}.\end{equation} The relevant components of the canonical
momentum of the $k^{th}$ massive neutrino in Eq.(\ref{p})
are\begin{eqnarray}p_{t}^{(k)}&=&p_{0}^{(k)}=m_{k}g_{00}\dot{t}+m_{k}g_{03}\dot{\varphi};\nonumber\\
p_{r}^{(k)}&=&m_{k}g_{11}\dot{r};\nonumber\\p_{\varphi}^{(k)}&=&m_{k}g_{33}\dot{\varphi}+m_{k}g_{03}\dot{t},\label{p0}\end{eqnarray}
where
$\dot{t}=\frac{dt}{ds},\dot{r}=\frac{dr}{ds},\dot{\varphi}=\frac{d\varphi}{ds}$.
Because the metric tensor components do not depend on the coordinate
$t$ and $\varphi$, their canonical momenta $p_{t}^{(k)}$ and
$p_{\varphi}^{(k)}$ are constant along the trajectory. In fact, the
momentum $p_{0}^{(k)}$conjugate to $t$ is the asymptotic energy of
the neutrino at $r=\infty$. It is stressed that it is the covariant
energy $p_{0}$ (not $p^{0}$) the constant of motion. Otherwise the
ambiguous definition of the energy will lead to the confusion in
understanding the neutrino oscillation.

The phase along the null geodesic from point A to point B is given
by \cite{null,zhangcm,stodolsky}
\begin{eqnarray}\Phi_{k}^{null}&=&\int_{A}^{B}p_{\mu}^{(k)}dx^{\mu}
=\int_{A}^{B}(p_{0}^{(k)}dt+p_{\varphi}^{(k)}d\varphi+p_{r}^{(k)}dr)
\nonumber\\&=&\int_{A}^{B}(p_{0}^{(k)}\frac{dt}{dr}+p_{\varphi}^{(k)}\frac{d\varphi}{dr}+p_{r}^{(k)})dr.\label{nu}\end{eqnarray}
We can obtain the following relations which are useful in the
calculation\begin{eqnarray}g^{00}&=&-\frac{g_{33}}{\Delta
sin^{2}\theta},g^{33}=-\frac{g_{00}}{\Delta
sin^{2}\theta},\nonumber\\g^{03}&=&\frac{g_{03}}{\Delta
sin^{2}\theta},g^{2}_{03}-g_{00}g_{33}=\Delta
sin^{2}\theta,\label{g00}\end{eqnarray}where
$\Delta=r^{2}-2Mr+a^{2}+Q^{2}$. Solving the equation (\ref{p0}) for
$\dot{t}$ and $\dot{\varphi}$, we
obtain\begin{eqnarray}\dot{t}=-\frac{g_{33}E_{k}+g_{03}L_{k}}{\Delta
sin^{2}\theta_{0}},\dot{\varphi}=\frac{g_{03}E_{k}+g_{00}L_{k}}{\Delta
sin^{2}\theta_{0}},\label{t}\end{eqnarray}where
$E_{k}=\frac{p_{0}^{(k)}}{m_{k}}$ and
$L_{k}=-\frac{_{p_{\varphi}^{(k)}}}{m_{k}}$ are the energy and
angular momentum per unit mass, respectively.

In the standard treatment of the neutrino oscillation, the neutrino
is usually supposed to travel along the
null\cite{null,fuller,ponte,zuber,boehm,bahcall} . Following the
standard treatment, we will calculate the phase along the light-ray
trajectory from $A$ to $B$.

The lagrangian appropriate to motions in the plane (for which
$\dot{\theta}=0$ and $\theta=$ a constant $=\theta_{0}$)
is\cite{chand}\begin{equation}\mathscr{L}=\frac{1}{2}(g_{00}\dot{t}^{2}+2g_{03}\dot{t}\dot{\varphi}+g_{11}\dot{r}+g_{33}\dot{\varphi}^{2}).\end{equation}
The Hamiltonian is given
by\begin{equation}\mathscr{H}=E_{k}\dot{t}-L_{k}\dot{\varphi}+\frac{p_{r}^{(k)}}{m_{k}}\dot{r}-\mathscr{L}.\end{equation}
Because of the independence of the Hamiltonian on $t$, we can deduce
that\begin{equation}2\mathscr{H}=E_{k}\dot{t}-L_{k}\dot{\varphi}+\frac{p_{r}^{(k)}}{m_{k}}\dot{r}=\delta_{1}=constant.\label{delta}\end{equation}
Without loss generality, we can set, $\delta_{1}=1$ for time-like
geodesics, $\delta_{1}=0$ for null geodesics. Substituting (\ref{t})
into (\ref{delta}) and setting $\delta_{1}=0$ for null geodesics, we
have the radial equation
\begin{equation}g_{11}\dot{r}^{2}=\frac{g_{33}E^{2}_{k}+2g_{03}E_{k}L_{k}+g_{00}L^{2}_{k}}{
sin^{2}\theta_{0}\Delta}.\label{g11}\end{equation} We define a new
function\begin{equation}V(r)=g_{33}E^{2}_{k}+2g_{03}E_{k}L_{k}+g_{00}L^{2}_{k}.\end{equation}The
different $V(r)$ determines the phase of the different trajectory.
From (\ref{g11}), we
get\begin{equation}\dot{r}=\frac{\sqrt{-V}}{\rho
sin\theta_{0}},\end{equation}where
$\rho^{2}=r^{2}+a^{2}cos^{2}\theta_{0}$. So, the equations governing
$t$ and $\varphi$
are\begin{equation}\frac{dt}{dr}=-\frac{\rho(g_{33}E_{k}+g_{03}L_{k})}{\Delta
sin\theta_{0}\sqrt{-V}},\frac{d\varphi}{dr}=\frac{\rho(g_{03}E_{k}+g_{00}L_{k})}{\Delta
sin\theta_{0}\sqrt{-V}}.\label{dtdr}\end{equation}

The mass-shell condition
is\cite{null}\begin{equation}m^{2}_{k}=g_{\mu\nu}p^{(k)\mu}p^{(k)\nu}=p_{\mu}^{(k)}p^{(k)\mu}=p_{0}^{(k)}p^{(k)0}
+p_{\varphi}^{(k)}p^{(k)\varphi}+p_{r}^{(k)}p^{(k)r}.\label{mass}\end{equation}
Substituting
$p^{(k)0}=g^{00}p_{0}^{(k)}+g^{03}p_{\varphi}^{(k)},p^{(k)\varphi}=g^{33}p_{\varphi}^{(k)}+g^{30}p_{0}^{(k)}$
and (\ref{p0}) into the equation of the mass-sell
condition(\ref{mass}), we
obtain\begin{equation}p^{(k)r}=\frac{m_{k}\sqrt{-V-
sin^{2}\theta_{0}\Delta}}{\rho sin\theta_{0}}.\end{equation}In the
process of calculation, the relations (\ref{g00}) are used. Applying
the relativistic condition $p_{0}^{k}\gg m_{k}$, we have the
relation\begin{equation}p^{(k)r}\simeq\frac{m_{k}}{\rho
sin\theta_{0}}(\sqrt{-V}-\frac{
sin^{2}\theta_{0}\Delta}{2\sqrt{-V}}).\end{equation} Adopting
(\ref{dtdr}) and $p^{(k)r}$, the phase along the null geodesics
(\ref{nu}) is approximated
by\begin{equation}\Phi^{null}_{k}\simeq\int_{A}^{B}\frac{m_{k}\rho
sin\theta_{0}dr}{2\sqrt{-V}}.\label{null}\end{equation} The phase
(\ref{null}) is a general result. The different function $V(r)$
corresponds to the different motion and determines the different
phase consequently. If $a=0$, we can obtain the oscillation phase in
the Reissner-Nordstorm space-time; if $Q=0$, the Kerr space-time
case is given. If $a=0,Q=0$, the function $V(r)$ reduces
to\begin{equation}V(r)=-r^{2}sin^{2}\theta_{0}E^{2}_{k}.\end{equation}The
phase (\ref{null}) becomes
to\begin{equation}\Phi^{null}_{k}=\int_{A}^{B}\frac{m^{2}_{k}}{2p_{0}^{(k)}}dr=\frac{m^{2}_{k}}{2p_{0}^{(k)}}(r_{B}-r_{A}).\end{equation}
This is just the phase in Schwarzschild
space-time\cite{fuller,null}.

The velocity of an extremely relativistic neutrino is nearly the
speed of light. In the standard treatment, the neutrino is supposed
to travel along the null
line\cite{ponte,null,fuller,zuber,boehm,bahcall}. Despite of this,
the propagation difference between a massive neutrino and a photon
can have important consequences and this tiny derivation becomes
important for the understanding of the neutrino oscillation.
Therefore, for more general situations, we start to calculate the
phase along the time-like geodesic. The factor of $2$ will be
obtained, when compared the time-like geodesic phase with the null
geodesic phase in the high energy limit. The classical orbit is
defined to a plane\cite{null,zhangcm}, $\theta=\theta_{0}$,
$d\theta=0$. The phase along the time-like geodesic
is\cite{zhangcm,12,ahlu,stodolsky}
\begin{equation}\Phi_{k}^{geod}=\int_{A}^{B}p_{\mu}^{(k)}dx^{\mu}
=\int_{A}^{B}(p_{0}^{(k)}\frac{dt}{dr}+p_{\varphi}^{(k)}\frac{d\varphi}{dr}+p_{r}^{(k)})dr.\end{equation}
For time-like geodesic, $\delta_{1}=1$, equation (\ref{delta})
becomes\begin{equation}E_{k}\dot{t}-L_{k}\dot{\varphi}+\frac{p_{r}^{(k)}}{m_{k}}\dot{r}=1,\label{28}\end{equation}
while the equations for $\dot{t}$ and $\dot{\varphi}$ (\ref{t}) are
the same for time-like geodesics\cite{chand}. Substituting $\dot{t}$
and $\dot{\varphi}$, we
have\begin{equation}\frac{ds}{dr}=\frac{1}{\dot{r}}=\frac{\sqrt{-g_{11}}}{\sqrt{-1-\frac{V}{\Delta
sin^{2}\theta_{0}}}}.\end{equation} So, we obtain the equations for
$dt/dr$ and $d\varphi/dr$ for time-like geodesics
\begin{eqnarray}\frac{dt}{dr}=-\frac{\sqrt{-g_{11}}}{\Delta\sin^{2}\theta_{0}}\frac{(g_{33}E_{k}+g_{03}L_{k})}{\sqrt{-1-\frac{V}{\Delta
sin^{2}\theta_{0}}}},\frac{d\varphi}{dr}=\frac{\sqrt{-g_{11}}}{\Delta\sin^{2}\theta_{0}}\frac{(g_{00}L_{k}+g_{03}E_{k})}{\sqrt{-1-\frac{V}{\Delta
sin^{2}\theta_{0}}}}.\end{eqnarray} According to mass shell
condition, $p_{r}^{(k)}$ is given by
\begin{equation}p_{r}^{(k)}=-m_{k}\sqrt{-g_{11}}\sqrt{-1-\frac{V}{\Delta\sin^{2}\theta_{0}}}.\end{equation}Thus, the phase along the time like
geodesic
is\begin{equation}\Phi_{k}^{geod}=\int_{A}^{B}\frac{m_{k}\sqrt{-g_{11}}dr}{\sqrt{-1-\frac{V}{\Delta\sin^{2}\theta_{0}}}}.\label{geod}\end{equation}
If the high energy limit is taken into account, Eq. (\ref{geod})
reduces
to\begin{equation}\Phi_{k}^{geod}\simeq\int_{A}^{B}\frac{m_{k}\rho\sin\theta_{0}dr}{\sqrt{-V}}=2\Phi_{k}^{null}.\end{equation}
It is often noted that the factor $2$ of the neutrino phase
calculations exists in the flat space-time\cite{lipkin,lipkin2} and
in the Schwarzschild space-time\cite{zhangcm,zhangcm4,12}, which is
believed to be the difference between the null geodesic and the time
like geodesic. The neutrino phase induced by the null condition, as
in the standard treatment, comes from the 4-momentum $p^{\mu}$
defined along the time-like geodesic, and the equation (\ref{g11})
governing $\dot{r}$ to the null geodesic. If the 4-momentum defined
along the null geodesic was instead used to compute the null phase,
we would obtain zero because of the null condition. When we
calculate the phase along the time-like geodesic, $\dot{r}$ in
(\ref{28}) is defined to the time-like geodesic. It is the
difference producing the factor $2$. It can be proved that the
neutrino phase along the null is the half of the value along the
time like geodesic in the high energy limit in a general curved
space-time(see APPENDIX A in literature\cite{zhangcm}).
\section{three special cases}
\subsection{Oscillation phases along the geodesics with
$L_{k}=aE_{k}$ in the equatorial plane} It is very important that
the geodesic is described in the equatorial plane $\theta=\pi/2$ in
the K-N space-time. The geodesics with $L_{k}=aE_{k}$ play the same
roles as the radial geodesics in the Schwarzschild and in the
Reissner-Nordstrom geometry. In this case, for null geodesic
$\dot{t}$, $\dot{\varphi}$ and $\dot{r}$ reduce to
\begin{equation}\dot{t}=\frac{r^{2}+a^{2}}{\Delta}E_{k};\dot{\varphi}=\frac{a}{\Delta}E_{k};\dot{r}=\pm E_{k}.\end{equation}
These equations in fact define the shear-free null-congruences which
we use for constructing a null basis for a description of the K-N
space-time in a Newman-Penrose formalism\cite{chand}. The function
$V(r)$ for null geodesic becomes to,
\begin{equation}V(r)=-r^{2}E^{2}_{k}.\end{equation}So, the phase along
the null
is\begin{equation}\Phi^{null}_{k}\simeq\int_{A}^{B}\frac{m_{k}\rho
sin\theta_{0}dr}{2\sqrt{-V}}=\int_{A}^{B}\frac{m_{k}}{2E_{k}}dr=\frac{m^{2}_{k}}{2p_{0}^{(k)}}(r_{B}-r_{A}),\end{equation}
which appears the same form as that of the Schwarzschild space-time
radial oscillation case.

We now turn to a consideration of the time-like geodesic case. The
equations for $\dot{t},\dot{\varphi}$ are the same as for the null
geodesics, while $\dot{r}$ becomes to
\begin{equation}\dot{r}=\sqrt{E_{k}^{2}+\frac{1}{g_{11}}}.\end{equation}Substituting $L_{k}=aE_{k}$ into
(\ref{geod}), we obtain the phase along the time-like
geodesic\begin{equation}\Phi_{k}^{geod}=\int_{A}^{B}\frac{m_{k}dr}{[(\frac{p_{0}^{(k)}}{m_{k}})^{2}+\frac{1}{g_{11}}]^{1/2}}.\label{x}\end{equation}
Compared with the phase along the radial time-like geodesic in the
Schwarzschild
space-time\cite{zhangcm},\begin{equation}\Phi_{k}^{geod}(Sch)=\int_{A}^{B}\frac{m_{k}dr}{\sqrt{(\frac{p_{0}^{(k)}}{m_{k}})^{2}-g_{00}}}
=\int_{A}^{B}\frac{m_{k}dr}{\sqrt{(\frac{p_{0}^{(k)}}{m_{k}})^{2}+\frac{1}{g_{11}}}},\end{equation}
we find that the oscillation phase with $L_{k}=aE_{k}$ in K-N
space-time has the similar form as the phase along the radial in
Schwarzschild space-time. Substituting
$g_{11}=-\frac{r^{2}}{\Delta}$ into equation (\ref{x}), we have
\begin{equation}\Phi_{k}^{geod}=\int_{A}^{B}\frac{m_{k}dr}{\sqrt{b+\frac{2M}{r}-\frac{a^{2}+Q^{2}}{r^{2}}}},\label{y}\end{equation}
where $b=(\frac{p_{0}^{(k)}}{m_{k}})^{2}-1$. Equation (\ref{y})can
be integrated directly to give
\begin{eqnarray}\Phi_{k}^{geod}&=&\frac{m_{k}}{b}\sqrt{br^{2}_{B}+2Mr_{B}-a^{2}-Q^{2}}-\frac{m_{k}}{b}\sqrt{br^{2}_{A}+2Mr_{A}-a^{2}-Q^{2}}
\nonumber\\&-&\frac{M
m_{k}}{b^{3/2}}ln\frac{br_{B}+M+\sqrt{b(br^{2}_{B}+2Mr_{B}-a^{2}-Q^{2})}}{br_{A}+M+\sqrt{b(br^{2}_{A}+2Mr_{A}-a^{2}-Q^{2})}}.\label{l=0}\end{eqnarray}
Eq.(\ref{l=0}) shows the effects of rotation $a^{2}$ on the
oscillation phase.

If $a=0$, we can obtain $\dot{t},\dot{\varphi},\dot{r}$ along the
radial null-geodesics in the equatorial plane in Reissner-Nordstrom
space-time\begin{equation}\dot{t}=\frac{r^{2}}{r^{2}-2Mr+Q^{2}},\dot{\varphi}=0,\dot{r}=\pm
E.\end{equation} Therefore, the phases along the radial null and
time-like geodesic in Reissner-Nordstrom space-time are given by,
respectively\begin{eqnarray}\Phi^{null}_{k}(RN)&=&\frac{m^{2}_{k}}{2p_{0}^{(k)}}(r_{B}-r_{A}),
\nonumber\\\Phi_{k}^{geod}(RN)&=&\int_{A}^{B}\frac{m_{k}dr}{\sqrt{b+\frac{2M}{r}-\frac{Q^{2}}{r^{2}}}}.\label{rn}\end{eqnarray}
Letting $a=0$ in (\ref{l=0}), the integral of equation (\ref{rn}) is
given.
\subsection{Oscillation phases along the geodesics with $L=0$ in the equatorial plane}
The geodesics with $L_{k}=0$ is another
important class of geodesics in K-N space-time. If the coordinate
$t$ and $\varphi$ has a relation $d\varphi/dt=-g_{03}/g_{33}$, the
canonical momentum $p_{\varphi}^{(k)}$ in (\ref{p0}) vanishes. The
corresponding $\dot{t}$, $\dot{\varphi}$ and $\dot{r}$ for null
geodesic are
\begin{equation}\dot{t}=-\frac{g_{33}}{\Delta}E_{k};\dot{\varphi}=\frac{g_{03}}{\Delta}E_{k};\dot{r}=\frac{\sqrt{-g_{33}}}{r}E_{k}.\end{equation}
And $\dot{r}$ for time-like geodesics is
\begin{equation}\dot{r}=\sqrt{\frac{1+g_{33}E^{2}_{k}/\Delta}{g_{11}}}.\end{equation}Substituting $L_{k}=0$ into (\ref{null})
and (\ref{geod}), the phases along the null and time-like geodesic
are given by, respectively
\begin{eqnarray}\Phi^{null}_{k}&=&\int_{A}^{B}\frac{m^{2}_{k}}{2p_{0}^{(k)}}\frac{r dr}{\sqrt{-g_{33}}}
=\int_{A}^{B}\frac{m^{2}_{k}}{2p_{0}^{(k)}}\sqrt{-g_{11}\widetilde{g_{00}}}dr,\label{41}\\\Phi_{k}^{geod}
&=&\int_{A}^{B}\frac{\sqrt{-g_{11}\widetilde{g_{00}}}m_{k}dr}{\sqrt{(\frac{p_{0}^{(k)}}{m_{k}})^{2}-\widetilde{g_{00}}}}.
\label{42}\end{eqnarray} where
$\widetilde{g_{00}}=g_{00}-g_{03}^{2}/g_{33}$. It is difficult to
integrate (\ref{41}) and (\ref{42}) directly. We can work out them
by expanding as $a^{2}$ when $a^{2}$ is a small quantity.
\subsection{Oscillation phase along the radial geodesic at $\theta=0$}
Unlike in the Schwarzschild and in the Reissner-Nordstrom
space-time, the event horizon does not coincide with the ergosphere
where $g_{00}$ vanishes in K-N space-time. This is an important
feature which distinguishes the K-N space-time from the others. The
ergosphere that is a stationary limit surface coincides with the
event horizon only at the poles $\theta=0$ and $\theta=\pi$. The
phase along the null geodesic in the direction $\theta=0$ can be
written
as\begin{eqnarray}\Phi_{k}^{null}&=&\int_{A}^{B}\frac{m_{k}\rho\sin\theta_{0}dr}{2\sqrt{-V}}
\nonumber\\&=&\int_{A}^{B}\frac{m_{k}\rho\sin\theta_{0}dr}{2\sqrt{r^{2}+a^{2}+\frac{a^{2}}{\rho^{2}}(2Mr-Q^{2})\sin\theta^{2}_{0}}\sin\theta_{0}}.
\label{0}\end{eqnarray} Substituting $\theta_{0}=0$, the equation
(\ref{0})
becomes\begin{equation}\Phi_{k}^{null}=\int_{A}^{B}\frac{m^{2}_{k}}{2p_{0}^{(k)}}dr=\frac{m^{2}_{k}}{2p_{0}^{(k)}}(r_{B}-r_{A}).\end{equation}
By similar calculation, the phase along the time-like geodesics at
$\theta=0$ is given
by\begin{equation}\Phi_{k}^{geod}=\int_{A}^{B}\frac{m_{k}dr}{(b+\frac{2Mr-Q^{2}}{r^{2}+a^{2}})^{1/2}},\end{equation}
where $b=(\frac{p_{0}^{(k)}}{m_{k}})^{2}-1$.
\section{proper oscillation length}
 The propagation of a neutrino is over its proper distance , but
$dr$ in (\ref{null}) is only a coordinate. The proper distance can
be written
as\cite{landau}\begin{equation}dl=\sqrt{(\frac{g_{0\mu}g_{0\nu}}{g_{00}}-g_{\mu\nu})dx^{\mu}dx^{\nu}}.\end{equation}
In K-N space-time, we
have\begin{equation}dl=\sqrt{-g_{11}dr^{2}+(\frac{g_{03}^{2}}{g_{00}}-g_{33})d\varphi^{2}}.\end{equation}
Substituting $\frac{d\varphi}{dr}$, we
obtain\begin{equation}dr=\frac{\sqrt{-g_{00}V}}{\sqrt{\Delta}E_{k}
sin\theta_{0}}dl.\label{drdl}\end{equation} In order to discuss
conveniently, we adopt the differential form of
(\ref{null})\begin{equation}d\Phi_{k}^{null}=\frac{m_{k}\rho
sin\theta_{0}dr}{2\sqrt{-V}}.\end{equation} Substituting
(\ref{drdl}), we have
\begin{equation}d\Phi_{k}^{null}=\frac{m_{k}^{2}}{2p_{0}^{(k)}}\sqrt{g_{00}}dl.\end{equation} It is assumed that
the mass eigenstates are taken to be the energy eigenstates, with a
common energy in the standard treatment. The equal energy assumption
is considered to be correct by some
authors\cite{grossman,lipkin,stodolsky2} and studied carefully in
papers\cite{zhangcm4,giunti2,leo}. In addition, it is adopted widely
in many literatures, for example\cite{null,fuller,zhangcm,giunti}.
$p_{0}$ will represent the common energy of different mass
eigenstates. In fact, the condition of equal momentum is also
adopted to study the neutrino oscillation. In the flat space-time,
both conditions (the equal energy and the equal momentum) present
practically the same neutrino oscillation results\cite{zhangcm4}.
There are conditions of time translation invariance and space
translation invariance in the flat space-time. So, energy
conservation and momentum conservation of a free particle are right.
In the curved (stationary) space-time, the energy of a particle is
conserved along the geodesic due to the existence of a time-like
killing vector field. However, the canonical conjugate momentum to
$r$, $p_{r}$ is not conserved because $(\frac{\partial}{\partial
r})^{a}$ is not killing in the curved (stationary) space-time.
Consequently, it is very difficult to study neutrino oscillation if
the condition of equal momentum is adopted in curved space-time. In
this section, our discussion is on the base of the results in the
standard treatment which the phase is calculated along the null.
Then, the phase shift which determines the oscillation
is\begin{equation}d\Phi^{null}_{kj}=d\Phi^{null}_{k}-d\Phi^{null}_{j}=\frac{\Delta
m_{k}^{2}}{2p_{0}}\sqrt{g_{00}}dl,\label{*}\end{equation} where
$\Delta m^{2}_{kj}=m^{2}_{k}-m^{2}_{j}$. The equation (\ref{*}) can
be rewritten
as\begin{equation}\frac{dl}{d(\frac{\Phi^{null}_{kj}}{2\pi})}=\frac{4\pi
p_{0}}{\Delta m^{2}_{kj}}\frac{1}{\sqrt{g_{00}}}=\frac{4\pi
p_{0}^{loc}}{\Delta m^{2}_{kj}}.\label{length}\end{equation} The
term $\frac{4\pi p_{0}}{\Delta m^{2}_{kj}\sqrt{g_{00}}}$ in
(\ref{length}) can be interpreted as oscillation length $L_{OSC}$
(which is defined by the proper distance as the phase shift
$\Phi^{null}_{kj}$ changing $2\pi$) measured by the observer at rest
at a position $r$, and $p_{0}^{loc}=p_{0}/\sqrt{g_{00}}$ is the
local energy. As $r\rightarrow\infty$, $p_{0}^{loc}$ approaches to
the energy $p_{0}$ measured by the observer at infinity. $\frac{4\pi
p_{0}}{\Delta m^{2}_{kj}}$ is the oscillation length in the flat
space-time. Equation (\ref{length}) is universal significance in
curved space-time. In fact, $\sqrt{g_{00}}$ is the gravitational red
shift factor which shows the effect of the gravitational field on
the oscillation length. Consider two static observers $O$ (the
radial coordinate $r$) and $O'$ (the radial coordinate $r'$). The
oscillation length measured by $O$ and by $O'$ is, respectively
\begin{equation}L_{OSC}(r)=\frac{4\pi p_{0}}{\Delta
m^{2}_{kj}}\frac{1}{\sqrt{g_{00}(r)}},\\L_{OSC}(r')=\frac{4\pi
p_{0}}{\Delta m^{2}_{kj}}\frac{1}{\sqrt{g_{00}(r')}}.\end{equation}
We can obtain the relation
\begin{equation}\frac{L_{OSC}(r')}{L_{OSC}(r)}=\frac{\sqrt{g_{00}(r)}}{\sqrt{g_{00}(r')}}.\end{equation}
If $r'>r$, we have $L_{OSC}(r')<L_{OSC}(r)$ and blue shift occurs.
Physically, the oscillation length is proportional to the local
energy of the neutrino. When the neutrino travels out of the
gravitational field, the local energy decreases. Consequently, the
neutrino oscillation length decreases and blue shift takes place.
From equation (\ref{length}), the oscillation length increases in
the gravitation field because of $0<g_{00}<1$ out of the the
infinite red shift surface. The effect of the gravitational blue
shift on the oscillation length may have the possible observable
effect from experiments. In the Schwarzschild space-time,
$g_{00}=1-2M/r$, we have
\begin{equation}L_{OSC}(Sch)=\frac{4\pi p_{0}}{\Delta m^{2}_{kj}}\frac{1}{\sqrt{1-2M/r}}.\end{equation}

In order to study the influence of Charge on the neutrino
oscillation, we consider the oscillation length in the
Reissner-Nordstrom space-time
\begin{equation}L_{OSC}(RN)=\frac{4\pi
p_{0}}{\Delta m^{2}_{kj}}\frac{1}{\sqrt{g_{00}}}=\frac{4\pi
p_{0}}{\Delta
m^{2}_{kj}}\frac{1}{\sqrt{1-\frac{2M}{r}+\frac{Q^{2}}{r^{2}}}}.\end{equation}
Compared with the case in the Schwarzschild space-time, the
oscillation length decreases due to the influence of charge $Q$.

The metric component $g_{00}$ in the K-N space-time
is\begin{equation}g_{00}=1-\frac{2Mr-Q^{2}}{\rho^{2}},\end{equation}where
$\rho^{2}=r^{2}+a^{2}\cos^{2}\theta$. In the equatorial plane, there
is, $g_{00}=1-2M/r+Q^{2}/r^{2}$, which is the same as $g_{00}$ in
the Reissner-Nordstrom space-time. Thus, it is concluded that the
neutrino oscillation length along the geodesics in the equatorial
plane in the K-N space-time is identical to that in the
Reissner-Nordstrom space-time and the rotating parameter $a^{2}$
does not work in this plane. Therefore, we have to select other
plane $\theta=\theta_{0}\neq\pi/2$ to highlight the effect of
rotation on the oscillation length. In the plane
$\theta=\theta_{0}$, the oscillation length can be written as
\begin{equation}L_{OSC}(K-N)=\frac{4\pi
p_{0}}{\Delta
m^{2}_{kj}}\frac{1}{\sqrt{1-\frac{2Mr-Q^{2}}{r^{2}+a^{2}\cos^{2}\theta_{0}}}}\label{62}\end{equation}
It is obvious that the oscillation length decreases too because of
the rotation of the gravitational field compared with that in R-N
space-time. Letting $Q=0$ in (\ref{62}), the oscillation length in
Kerr space-time is given by
\begin{equation}L_{OSC}(Kerr)=\frac{4\pi
p_{0}}{\Delta
m^{2}_{kj}}\frac{1}{\sqrt{1-\frac{2Mr}{r^{2}+a^{2}\cos^{2}\theta_{0}}}}.\end{equation}
Comparing with (\ref{62}), the charge $Q$ shortens the oscillation
length. We can obtain that the oscillation length varies with
$\theta_{0}$
by\begin{equation}\frac{d}{d\theta_{0}}L_{OSC}(K-N)=\frac{4\pi
p_{0}}{\Delta
m^{2}_{kj}}(g_{00})^{3/2}\frac{2Mr-Q^{2}}{(r^{2}+a^{2}\cos^{2}\theta_{0})^{2}}a^{2}\sin\theta_{0}\cos\theta_{0}.\end{equation}
In K-N space-time, we conclude that the oscillation length increases
with $\theta$ within $0<\theta<\pi/2$, and it becomes maximum in the
equatorial plane. Then, it decreases with $\theta$ within
$\pi/2<\theta<\pi$. At the direction $\theta=0$ and $\theta=\pi$,
the oscillation length occurs minimum,
\begin{equation}L_{OSC}(K-N)=\frac{4\pi
p_{0}}{\Delta
m^{2}_{kj}}\frac{1}{\sqrt{1-\frac{2Mr-Q^{2}}{r^{2}+a^{2}}}}.\end{equation}

In summary, the gravitational field lengthens oscillation length;
both the rotation $a^{2}$ and the charge $Q$ shorten the oscillation
length.

\section{conclusion and discussion}
In this paper, we have given the phase of mass neutrino propagating
along the null and the time like geodesic in the gravitational field
of a rotating symmetric and charged object, which is described by
Kerr-Newman metric. Most astrophysical bodies in universe have
rotation and charge generally. Thus the work about the neutrino
oscillation in the K-N space time is important and meaningful for
the black hole astrophysics. We work out the general formula of
oscillation phase on the equal $\theta=\theta_{0}$ plane with the
generality. The phase along the geodesic is the double of that along
the null in the high energy limit, which is the same in the cases in
flat and Schwarzschild space-time. By setting $\theta=\pi/2$, the
phases in the equatorial plane are given. As $a=0$ or $Q=0$, we
obtain the phases in the R-N space-time or in the Kerr space-time.
Moreover, we study three special cases in K-N space-time: geodesics
with $L=aE$; geodesics with $L=0$; radial geodesics at $\theta=0$.
Among them, the geodesics with $L=aE$ have the same importance as
the radial geodesics in the Schwarzschild and in the R-N geometry.
The phases obtained are very similar in form to the cases along the
radial geodesics in the Schwarzschild and in the R-N space-time.

In Sec.5, the proper oscillation length in the K-N space time is
studied in detail. We find that oscillation length in curved
space-time is proportional to the local energy, which is regraded as
the neutrino "climbs out of the gravitational potential well". So,
the blue shift occurs. Then, the effects of rotation and charge of
the space-time on the oscillation length are given. Because of the
correction of the gravitation field, the oscillation length
increases, compared with the flat space time case. However, both the
rotation $a^{2}$ and the charge $Q$ shorten the oscillation length.
It is noted, the rotation has null contribution to the length in the
equatorial plane in K-N space-time, because red shift factor is
independence of $a^{2}$ in this plane. Finally, we remark that our
result exists generality, which  can be exploited  to study the
neutrino oscillation near the rotating compact star, neutron star or
black hole.

\end{document}